\documentclass{sf2a-conf}
\usepackage{graphicx}

\def\vsini{$V\!\sin i$}
\def\teff{T$_ {\rm{eff}}$}

\def\logg{log~{\it g}}

\begin{document}
\TitreGlobal{SF2A 2007}
%%-----------------------------
%%      the top matter
%%-----------------------------
\title{On the nature of early-type emission line objects in NGC6611}

% The preferred form for each name is: initial(s) of the forename(s) followed by the family name.
%If there is more than one author, the order is optional. . 
%If the authors have different affiliations, each  name has to be followed by \address. 
%Numbers referring to different addresses should be attached to each author, pointing 
% to the corresponding institute.

\author{C. Martayan$^{1,}$} \address{Royal Observatory of Belgium, 3 avenue circulaire, 1180 Brussels, Belgium}
\address{GEPI, Observatoire de Paris, CNRS, Universit\'e Paris Diderot; 5 place Jules Janssen, 92195 Meudon Cedex, France}
\author{M. Floquet$^2$}
\author{A.-M. Hubert$^2$}
\author{J. Fabregat}\address{Observatorio Astron\'omico de Valencia, edifici Instituts d'investigaci\'o, Poligon la Coma, 46980 Paterna Valencia, Spain} 
\author{Y. Fr\'emat$^1$}
\author{D. Baade}\address{European Organisation for Astronomical Research in the Southern Hemisphere, Karl-Schwarzschild-Str. 2, D-85748 Garching b. Muenchen, Germany} 
\author{C. Neiner$^2$}

% Define here your short title
\runningtitle{On the nature of early-type emission line objects in NGC6611}
% Do not worry about this

\setcounter{page}{1}

% Repeat the authors here, this will help to make the final index (Name first and Initial after)
\index{Martayan C.} 
\index{Floquet M.}
\index{Hubert A.-M.} 
\index{Fabregat J.} 
\index{Fr\'emat Y.}
\index{Baade D.}
\index{Neiner C.}

\maketitle

\begin{abstract}
The number and the nature of emission line stars in the young open
cluster NGC6611 is still the object of debates. Due to the presence of a
strong and variable nebulosity in the cluster, the number of emission line
stars is highly depending on the technique and the resolution used for the
observations. Thanks to observations with the ESO-WFI, in slitless
spectroscopic mode, and with the VLT-GIRAFFE we have been able to
disentangle the circumstellar and nebular emissions. We confirm the small
number of true emission line objects and we precise their nature: mainly
Herbig Be stars.
\end{abstract}
%
%%-----------------------------
%%      your text
%%-----------------------------
\section{Introduction}

NGC6611 is a young open cluster with log(age)~=~6.2 or 6.8, depending on the authors. According to
Hillenbrand et al. (1993) and de Winter et al. (1997) it contains a great number of emission line
stars (ELS), whereas Herbig \& Dahm (2001) only found a small number. It is however worth noticing
that the two first studies were carried out using slit spectrographs, while Herbig \& Dahm (2001) used a slitless
instrument not sensitive to the surrounding emission originating from the Eagle nebula.

In order to further investigate the occurence of emission line stars in NGC6611 and to determine
the nature of objects, we used the Wide Field Imager (WFI) at ESO in slitless spectroscopic mode associated 
to a 200~nm passband H$\alpha$ filter. We further also made use of the multi-object spectrograph GIRAFFE
at the VLT in MEDUSA mode. The spectra obtained in this way allowed us: 1) to determine the
stellar parameters, 2) to disentangle the circumstellar and nebular emissions, and 3) to determine the
true nature of the targets.

\section{Emission line stars}

\subsection{WFI observations}

With WFI, we obtained $\sim$15000 spectra of the sources in NGC6611 and its surrounding field.
Due to the fact that WFI in slitless mode is not sensitive to the ambient nebular emission, we 
listed the stars with and without circumstellar (CS) emission. However, this slitless mode does not allow
the detection of faint CS emissions. A small number of ELS was identified and preselected for the VLT-GIRAFFE observations.

\subsection{VLT-GIRAFFE observations}

\begin{figure*}[h!]
    \centering
    \resizebox{\hsize}{!}{\includegraphics[angle=0]{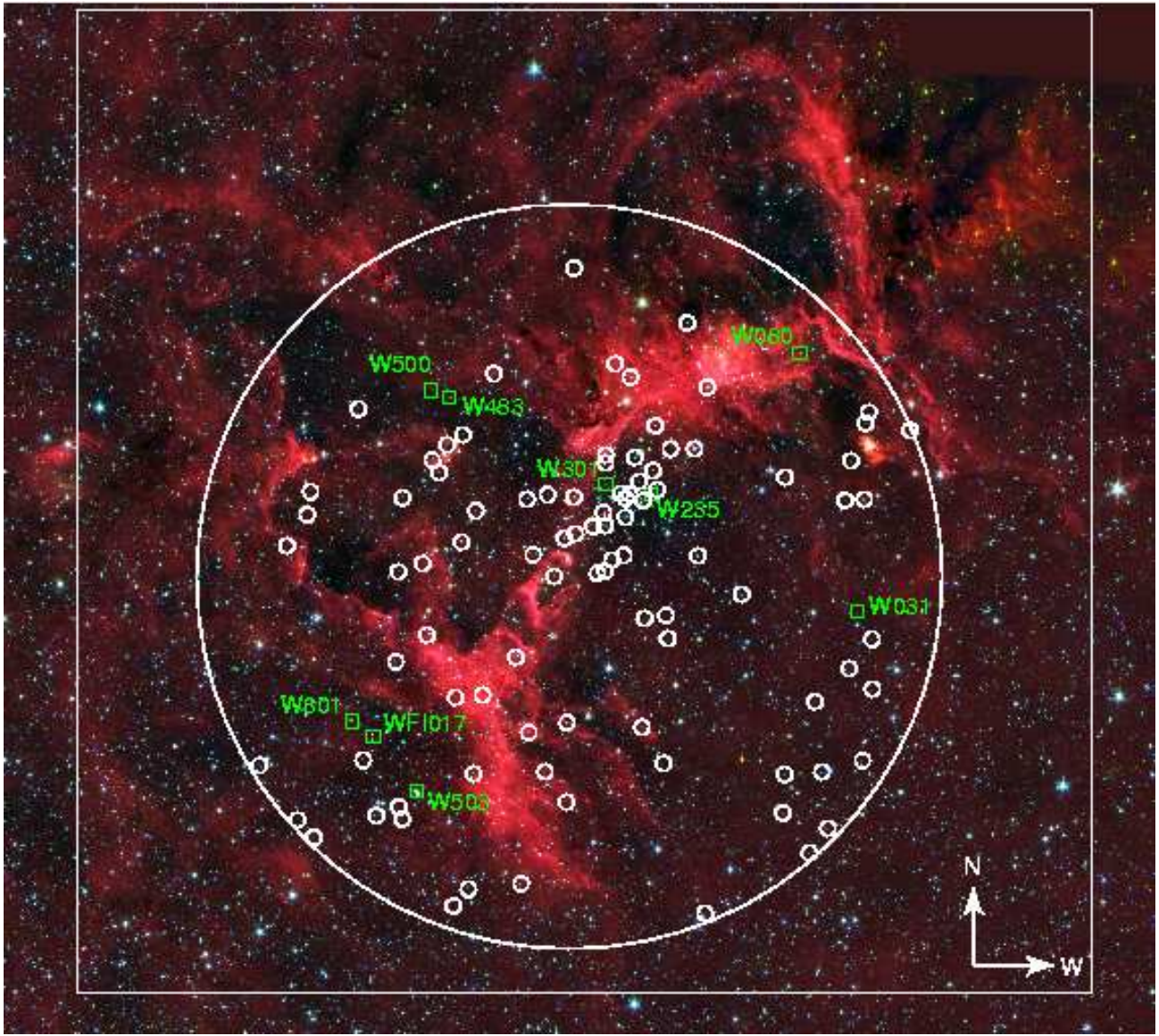}}
    \caption{Location of the stars observed with the VLT-GIRAFFE in the field of NGC6611. The image
    is a RGB mosaic with 3 channels of SPITZER (B: 3.6$\mu$m, G: 4.5$\mu$m, and R: 8$\mu$m).  The
    small white circles are for the non-ELS stars, the green boxes are for the ELS stars. The large
    white circle shows the field of GIRAFFE and the large white box shows the field of the
    WFI-spectro. WFI017 stands for WFI[N6611]017.}
    \label{loc6611}
\end{figure*}

The 100 objects finally observed with GIRAFFE are shown in Fig.~\ref{loc6611}. 

\begin{figure}[h]  
\centering 
\begin{tabular}{cc}
\includegraphics[width=8.2cm]{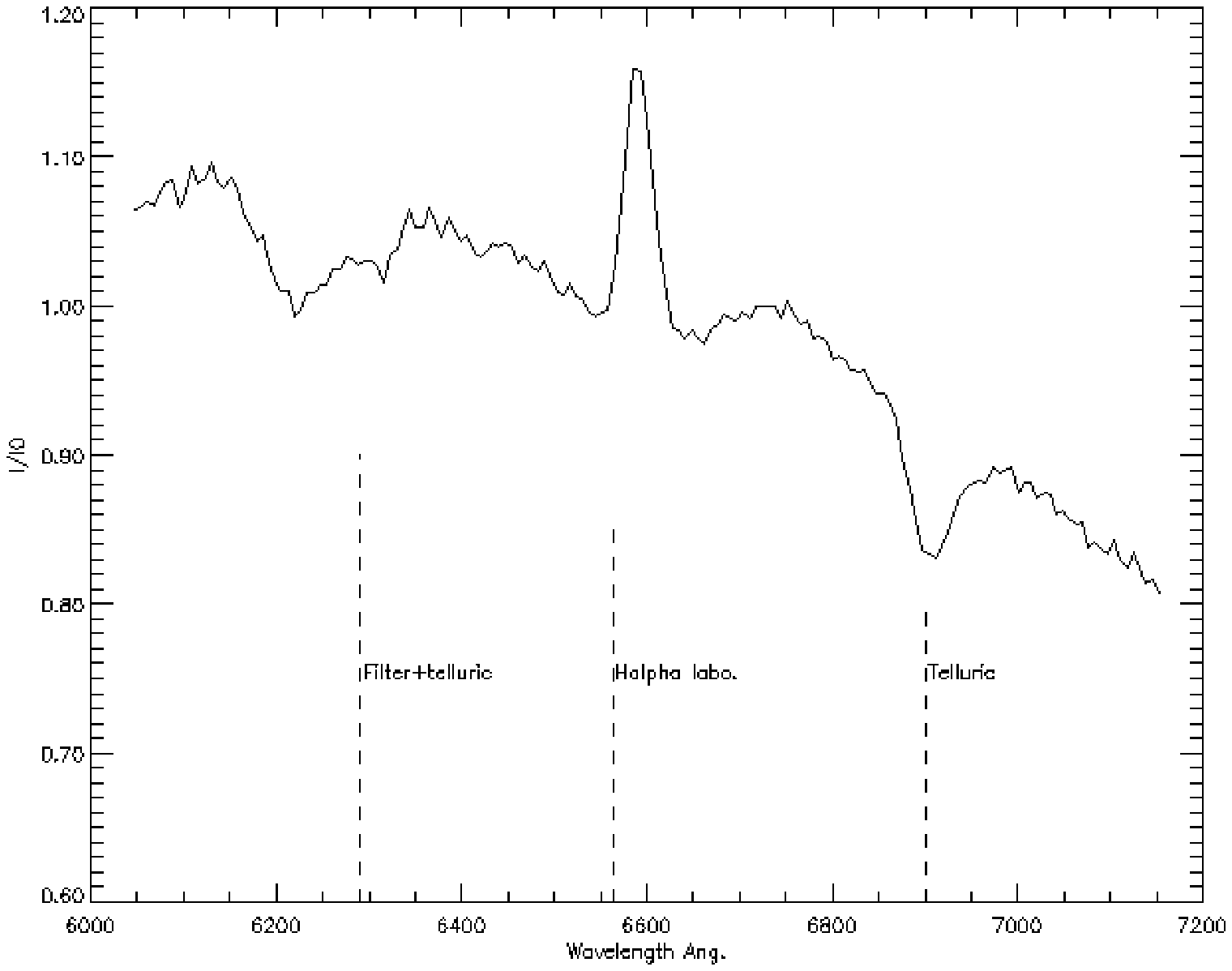}&\includegraphics[width=7cm]{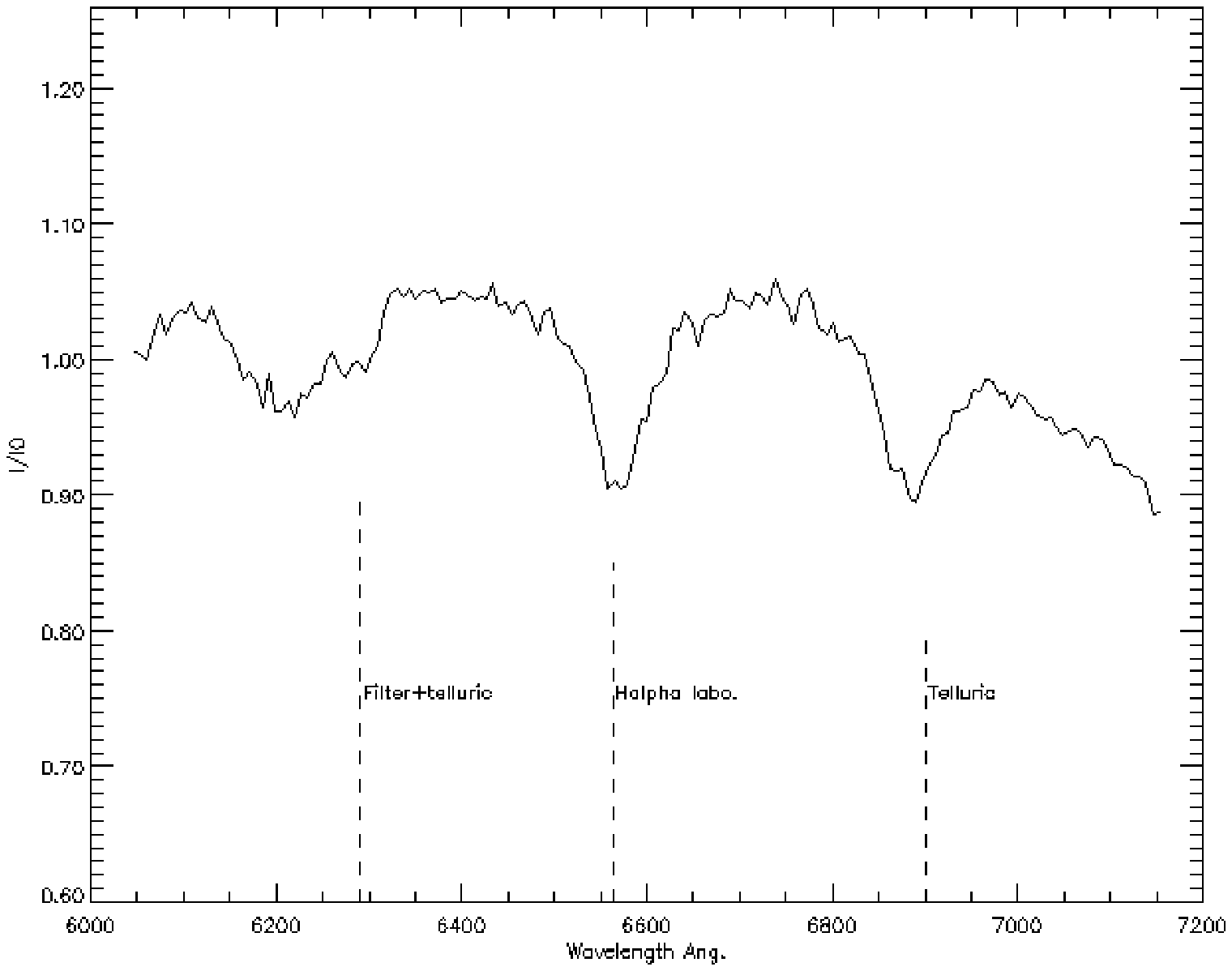}\\
\includegraphics[width=8cm]{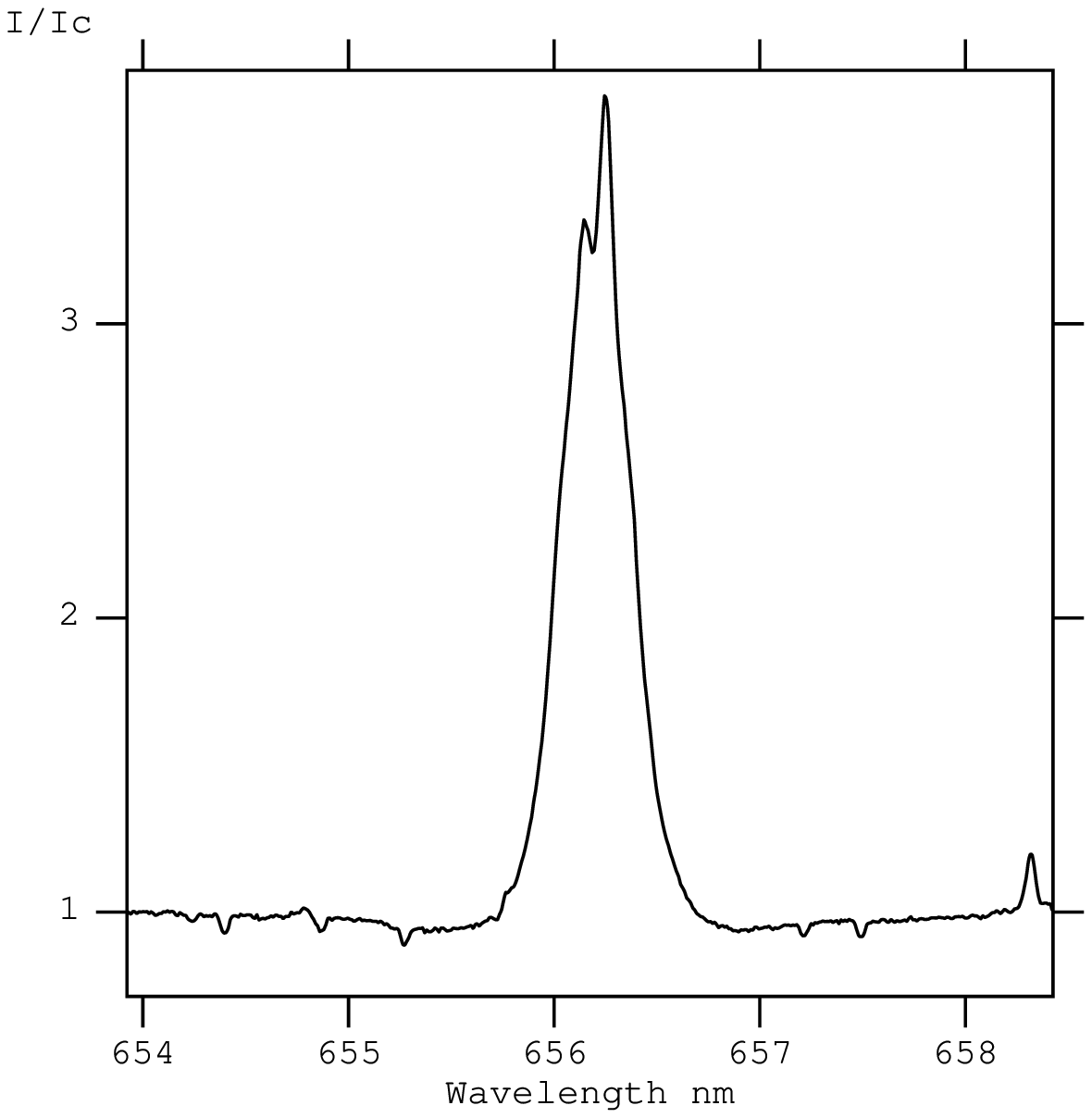}&\includegraphics[width=8cm]{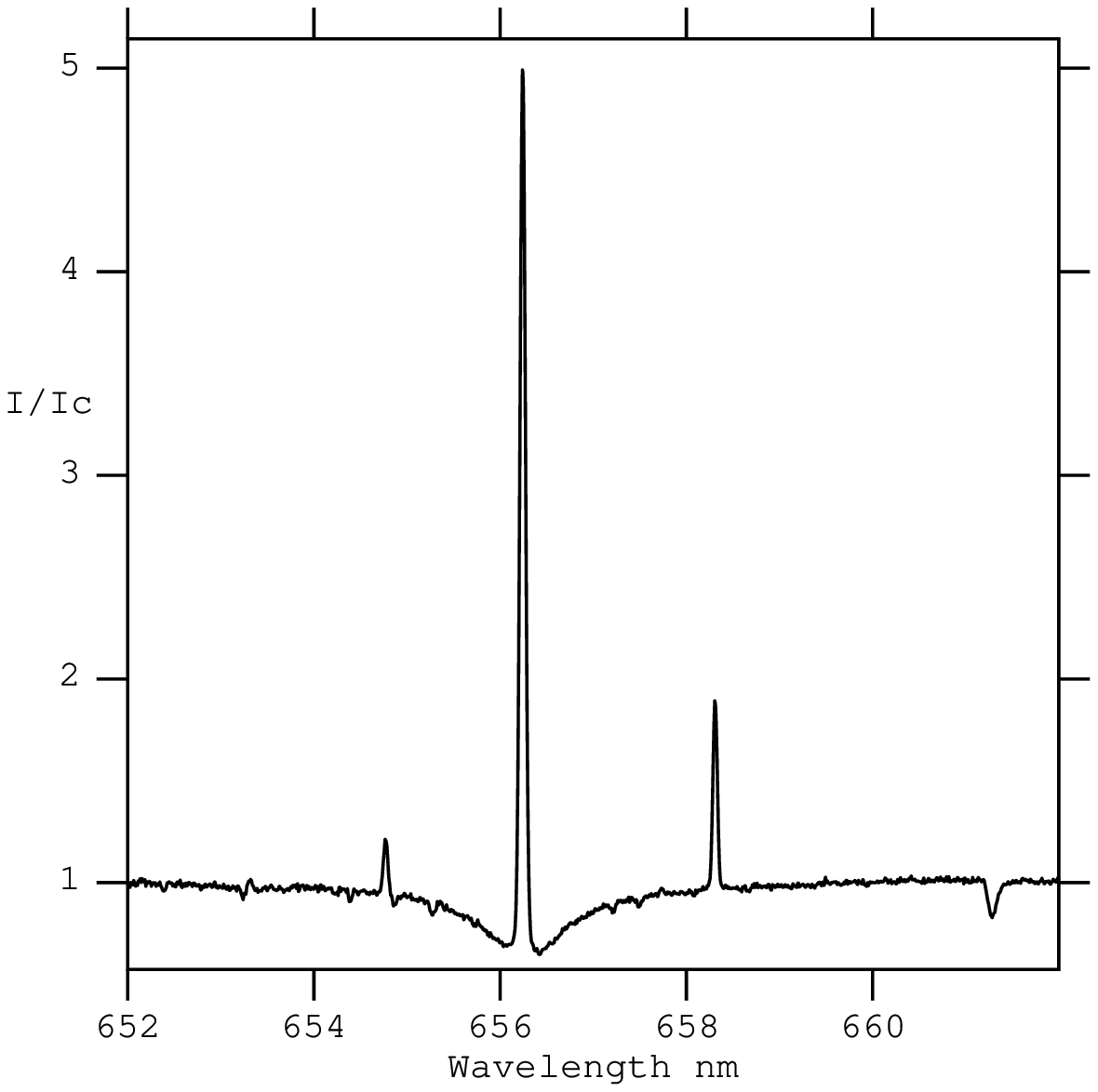}
\end{tabular}
\caption{Examples of WFI (upper panels) and GIRAFFE (lower panels) H$\alpha$ spectra. W483 (left panels) displays a CS H$alpha$ emission line while  W371 (right panels) does not show any emission in WFI spectra but a nebular contamination in GIRAFFE spectra.}
\label{wfi_spectra} 
\end{figure}

Among them, only 9 were identified as ``true'' CS ELS.
The main part of the previously known ELS had their spectra contaminated by nebular emission, as 
shown in Figure~\ref{wfi_spectra} where the H$\alpha$ regions obtained with WFI and GIRAFFE are displayed for two 
cases:  W483, a true ELS, and  W371, misidentified as ELS in previous slit spectroscopy.

\section{Nature of NGC6611 stars}
To investigate the nature of the stars  observed with GIRAFFE, we determined their fundamental parameters and studied their spectral energy distributions (SED).

\subsection{Fundamental parameters}
We determined the fundamental parameters (\teff, \logg~and \vsini) by fitting the GIRAFFE data with
synthetic spectra following a procedure described in Fr\'emat et al. (2006). Other parameters 
(mass, radius, age, luminosity) were then estimated by interpolation in theoretical evolutionary
tracks computed for a solar metallicity (Schaller et al. 1992). A part of stars in the sample (mainly the massive stars) 
are young and lie close to the ZAMS. However, the analysis of our results demonstrates
that a group of intermediate mass stars  are too old by comparison to the age of this star-formation region.
Consequently, these stars must be, in fact, considered as pre-main sequence stars (PMS), which go to reach the ZAMS. 
We therefore re-estimated their age using the PMS evolutionary tracks computed by Palla \& Stahler (1993) 
and Iben(1965).

\subsection{Infrared study}
The intrinsic interstellar reddening E(B-V) was measured for each star by means of the interstellar lines at 443.0 and 661.3 nm. It was used to correct the UBVI, JHK ( 2MASS survey), 3.6$\mu$m, 4.5$\mu$m,
5.7$\mu$m and 8$\mu$m (SPITZER) magnitudes. Each SED is normalized to the SED of a normal B star.
For certain of non ELS stars, as well as for 5 of the ELS, the SED show an infrared excess.
The origin of this more or less strong infrared excess could be a disk or cocoon around the central star.
Such an infrared excess is compatible with Herbig Ae/Be stars or with PMS stars.
This infrared study confirms the presence of PMS stars and allows to determine/confirm that the main part of the observed  ELS are Herbig Ae/Be stars. 

\subsection{Nature of the true ELS}
From the study of the H$\alpha$ emission line, the  fundamental parameters and  the infrared excess,
we concluded that: 
\begin{itemize}
\item the stars WFI[N6611]017, W080, W235, W483, W500, W503 are Herbig Ae/Be stars.
\item the star W031 is a possible Herbig Ae/Be star but a doubt remains.
\item the star W301 is a classical Be star.
\end{itemize}

\section{Conclusions}

Slitless spectroscopy (WFI) as well as high resolution slit spectroscopy (VLT-GIRAFFE) confirm there are only few true emission line stars in NGC6611  as shown by Herbig \& Dahm (2001).
Thanks to the investigation of age and IR excess combined with spectral information we identified 
 8 Herbig Ae/Be stars and 1 possible classical Be star as well as   several PMS stars without 
emission. Moreover, considering the more massive stars of the cluster which lie on the main sequence
we estimated the age of NGC6611; we found  log(age)=6.8 $\pm 0.2$.

\end{document}